\documentclass[12pt]{article}
\usepackage{a4,amsmath,epsfig,amssymb}
\usepackage[latin1]{inputenc}
\usepackage{graphicx}
\usepackage{caption}
\usepackage{subcaption}
\usepackage{feynmp}
\begin{document}
\bigskip\bigskip
\begin{center}
{\large \bf Searching for New Physics\\
in Semi-Leptonic Baryon Decays}
\end{center}
\vspace{8pt}
\begin{center}
\begin{large}

E.~Di Salvo$^{a,b,}$\footnote{Elvio.Disalvo@ge.infn.it},
Z.~J.~Ajaltouni$^{a,}$\footnote{ziad@clermont.in2p3.fr}
\end{large} 

\bigskip
$^a$ 
Laboratoire de Physique Corpusculaire de Clermont-Ferrand, \\
IN2P3/CNRS Universit\'e Blaise Pascal, 
F-63177 Aubi\`ere Cedex, France\\

\noindent  
$^b$ 
Dipartimento di Fisica 
 Universit\'a di Genova \\
and I.N.F.N. - Sez. Genova,\\
Via Dodecaneso, 33, 16146 Genova, Italy \\  

\noindent  

\vskip 1 true cm
\end{center} 
\vspace{1.0cm}

\vspace{6pt}
\begin{center}{\large \bf Abstract}

We propose two different and complementary observables for singling out possible signals of physics beyond the standard model in the semi-leptonic decays $\Lambda_b \to \Lambda_c \ell{\bar \nu}_{\ell}$, both with the $\tau$ lepton and with a light lepton. The two observables are the partial decay width and a T-odd asymmetry, whose respective sensitivities to scalar and/or pseudo-scalar coupling are calculated as functions of the parameters characterizing new physics. Two different form factors are used. Three particular cases are discussed and analyzed in detail.
\end{center}

\vspace{10pt}

\centerline{PACS numbers: 13.30.Ce, 12.15.-y, 12.60.-i}

\newpage

\section{Introduction}
The search for new physics (NP) beyond the standard model (SM), a major aim in high energy physics, is especially stimulating now, that the Higgs boson, the last ingredient of the SM, has been 
found\cite{cms,atl}. Indeed, the SM has achieved a resounding success, as a lot of data confirm its predictions and no unambiguous sign of NP has been found till now. However, that model presents
 several unsatisfactory aspects\cite{pkn,bwh}: it has to be regarded, at best, as an effective low-energy approximation\cite{alt} of the theory of the fundamental particles. Therefore, it is 
essential to realize experiments where, either a clear contradiction with the SM is found, or at least more stringent constraints on the physics beyond it are established.

Recently, tensions have been found between the SM predictions and the experimental values of some partial widths of semi-leptonic decays of $B$, like 
$B\to K^*\ell^+\ell^-$\cite{lhcb1,lhcb2} and 
$B \to D^{(*)} \ell \nu_{\ell}$\cite{bb1,bb2,bel,lhcb3}. However, such discrepancies are not definitive, owing to the experimental and theoretical errors, the latter ones being related to the 
transition form factors, which are affected by serious uncertainties\cite{as,dm}. Therefore it is necessary to get confirmations from other channels whose basic processes are the same as in the
 above mentioned $B$ decays, in particular, from $\Lambda_b$ decays to $\Lambda\ell^+\ell^-$\cite{dm,gu} and to $\Lambda_c\ell^-{\bar \nu}_{\ell}$\cite{swd,dut,gu1,hgi}. Indeed, the baryonic form 
factors may pose less difficulties than the corresponding mesonic ones\cite{dm}, in any case they are calculated to an increasing precision\cite{gu1,dlm,jjo}. 

Moreover, the above mentioned baryon decays involve spinning particles both in the initial and in the final state, which offers the opportunity of examining also another kind of observables, the T-odd 
asymmetries, based on triple product correlations. These were largely studied in the past decades\cite{va,gv,wf} and were recommended for searching for time reversal violations (TRV) and, possibly, 
for NP\cite{gv,bl,bdl,arg,gr,gr1}; see ref. \cite{ajd} for a more complete review. This method of analysis appears especially promising for the search for NP in semi-leptonic baryon decays, in that
 the T-odd asymmetry is vanishing according to the SM. Indeed, the final-state interactions, which cause biases to 
non-leptonic decays by fictitious T-odd interactions\cite{ajd}, are absent in this case. Therefore, a nonzero value of this observable would unambiguously indicate NP and direct TRV.

It is the aim of the present letter to calculate, in a model independent way, the sensitivities to NP of the partial width and of the asymmetry of the decays 
\begin{equation}
\Lambda_b \to \Lambda_c \ell^- \bar{\nu}_{\ell},  \label{slh} 
\end{equation}
both with the $\tau$ lepton and with a light lepton.
If the parent resonance is polarized, or the polarization of the final baryon (or of the lepton) is measured, a nonzero T-odd asymmetry may 
be defined, {\it i. e.}, 
\begin{equation}
{\cal A} = \frac{N^+ - N^-}{N^+ + N^-},   \label{as}
\end{equation}
where $N^{\pm}$ is the number of events for which the T-odd correlation  
\begin{equation}
C = {\bf p}_f \times {\bf p}_{\ell} \cdot {\bf P} \label{crr}
\end{equation}
is positive (negative). Here ${\bf p}_{f(\ell)}$ and ${\bf P}$ denote, respectively, the momentum of the final baryon (charged lepton) and the polarization vector, in the center-of-mass system of the initial baryon.
In the present paper, we limit ourselves to the case of a polarized $\Lambda_b$. For the asymmetry (\ref{as}) to be nonzero, the above decay must derive contributions from at least two amplitudes - the SM one and some NP amplitude -, endowed with a nontrivial relative phase\cite{va,ajd}; moreover, we shall establish that the NP coupling must be of the (pseudo-)scalar type.
 
We assume the effective NP hamiltonian proposed by other authors\cite{swd,dut,ddg}; however, for the sake of simplicity, we limit ourselves to analyzing the effects of a scalar and/or pseudo-scalar coupling. Indeed, as claimed above, the asymmetry is insensitive to the vector and to the axial coupling, so acting a selection of NP interactions\cite{ddg}; furthermore, the (pseudo)-scalar couplings seem to play an important role in possible NP of the $B^0_s\to J/\psi \Phi$ and $B\to\Phi K^0_s$ decays\footnote{ See refs. 14 of ref. \cite{ddg}}. 

The two observables mentioned above are expressed and studied as functions of parameters that characterize NP. As regards the baryonic current, we adopt prevalently the Isgur-Wise\cite{iw} (IW) model, quantitatively reliable\cite{jjo,yor,klw,gkl,kkk}: as we shall see, it reproduces quite well the experimental value of the partial width of the decay (\ref{slh}) with a light lepton. We compare the IW predictions with those deduced from one of the more sophisticated expressions of the hadronic current\cite{swd,dut,dlm1}, leaving a more complete discussion of this point to a future paper.

The letter is organized as follows. In sect. 2, we give the general formulae concerning the differential decay width. Sects. 3 and 4 are dedicated to illustrating the behaviors, respectively, of the 
partial decay width and of the T-odd asymmetry, as functions of the parameters that characterize NP. Lastly, we draw some conclusions in sect. 5. 

\section{General Formulae}
   
According to the IW assumption, the amplitude for the above mentioned decay reads as
\begin{equation}
{\cal M} = V_{cb}\zeta_0(q^2)(h_s J_{\mu} j^{\mu}+ e^{i\varphi} h_n J j).
\end{equation}
	
Here $\varphi$ is the relative phase of the NP amplitude to the SM one. Moreover,

\begin{equation}
h_s = \frac{g^2}{8M_W^2} = \frac{G}{\sqrt{2}}, ~~~~~ 
h_n = xh_s, \label{cnst}
\end{equation}
\begin{eqnarray}
j_{\mu} &=& {\bar u}_{\ell}\gamma_{\mu}(1-\gamma_5)v, ~~~~~ 
j = {\bar u}_{\ell}(1-\gamma_5)v, \label{lcur}
\\
J_{\mu} &=&  {\bar u}_f \gamma_{\mu}(1-\gamma_5) u_i,  ~~~~~ J = \frac{q^{\alpha}}{\delta m_Q}{\bar u}_f \gamma_{\alpha}(1+\rho\gamma_5) u_i,
\end{eqnarray}
\begin{equation}
q = p_i-p_f = p_{\ell}+p.
\end{equation}

The subscripts $i$, $f$ and $\ell$ refer, respectively, to the initial and final baryon and to the charged lepton, while $p$ ($v$) is the four-momentum (four-spinor) of the anti-neutrino. 
$\zeta_0(q^2)$ is
the form factor of the $\Lambda_b \to \Lambda_c$ transition in the IW model\cite{iw}. $\delta m_Q$ and $V_{cb}$ are, respectively, the difference between the masses of 
the active quarks ($b$ and $c$ in our case) and the CKM matrix element of that transition. $G$ is the Fermi constant and $h_{s(n)}$ expresses the SM (NP) strength. Lastly, $\rho$ is a 
complex parameter which determines the mixing of scalar and pseudo-scalar coupling. Our parameters are related in a simple way to the (complex) couplings of ref. \cite{swd}:
\begin{equation}
g_S = x e^{i\varphi},  ~~~~~ \ ~~~~~ \ g_P = \rho x e^{i\varphi}.
\end{equation}
In the present note, we limit ourselves to real values of $\rho$; in particular, we consider the following three limiting cases:

- if $\rho$ = 0, only the scalar coupling is considered\cite{swd};

- the case of $\rho$ $\to$ $\infty$, $x$ $\to$ 0, $x \rho$ $\to$ $z$, corresponds to a pure pseudo-scalar coupling term\cite{swd};

- the two-higgs doublet model is recovered for\cite{klld} 
\begin{equation}
\rho = \frac{m_b-m_c}{m_b+m_c} \sim 0.53. \label{2hdm}
\end{equation}

The observables that we study in this paper are derived from the differential decay width
\begin{equation}
d\Gamma = \frac{1}{2m_i} \sum|{\cal M}|^2 d\Phi, \label{ddw}
\end{equation}
where $d\Phi$ is the phase space and $\sum|{\cal M}|^2$ denotes average (sum) over the initial (final) polarizations of the particles involved. But we have
\begin{equation}
\sum |{\cal M}|^2 = |V_{cb}|^2 \frac{G^2}{2}\zeta_0^2(q^2)[H_{\mu\nu}\ell^{\mu\nu} +
\frac{2x}{\delta m_Q} \Re(\ell^{\mu} q^{\nu} I_{\mu\nu}e^{-i\varphi})+
x^2 K_{\mu\nu}\frac{q^{\mu} q^{\nu}}{(\delta m_Q)^2}\ell]. \label{modsq}
\end{equation}
Here
\begin{eqnarray}
\ell^{\mu\nu} &=& Tr[p\hspace{-0.45 em}/\gamma^{\nu}(1-\gamma_5)
(p\hspace{-0.45 em}/_{\ell}+m_{\ell})\gamma^{\mu}(1-\gamma_5)], ~~~~~ \ ~~~~~ \
\\
\ell^{\mu}~ &=& Tr[p\hspace{-0.45 em}/(1+\gamma_5)(p\hspace{-0.45 em}/_{\ell}+m_{\ell})
\gamma^{\mu}(1-\gamma_5)], ~~~~~ \ ~~~~~ \ ~~~~~ \
\\
\ell~~ &=& Tr[p\hspace{-0.45 em}/(1+\gamma_5)(p\hspace{-0.45 em}/_{\ell}+m_{\ell})
(1-\gamma_5)], ~~~~~ \ ~~~~~ \ ~~~~~ \ ~~~~~ \
\\
H_{\mu\nu} &=& 1/2Tr[(p\hspace{-0.45 em}/_i+m_i) (1+\gamma_5 s\hspace{-0.45 em}/) \gamma_{\nu}
(1-\gamma_5)(p\hspace{-0.45 em}/_f+m_f)\gamma_{\mu}(1-\gamma_5)],
\\ 
I_{\mu\nu} &=& 1/2 Tr[(p\hspace{-0.45 em}/_i+m_i) (1+\gamma_5 s\hspace{-0.45 em}/) \gamma_{\nu}(1+\rho\gamma_5)(p\hspace{-0.45 em}/_f+m_f)\gamma_{\mu}(1-\gamma_5)],
\\
K_{\mu\nu} &=& 1/2 Tr[(p\hspace{-0.45 em}/_i+m_i) (1+\gamma_5 s\hspace{-0.45 em}/) \gamma_{\mu}(1+\rho\gamma_5)(p\hspace{-0.45 em}/_f+m_f)\gamma_{\nu}(1+\rho\gamma_5)],
\end{eqnarray}
$s$ being the polarization four-vector of the $\Lambda_b$, $s$ $\equiv$ $(0, {\bf P})$
 in the $\Lambda_b$ rest frame, with $|{\bf P}|$ $\leq$ 1. Performing calculations and substituting into eq. (\ref{ddw}) yields 
\begin{equation}
d\Gamma = \frac{1}{2m_i} |V_{cb}|^2 \frac{G^2}{2}\zeta_0^2(q^2) 
[2^7 R_s + 2^5 x \frac{m_{\ell}}{\delta m_Q} (R_i^e cos\varphi+ R_i^o sin\varphi)
+2^4 \frac{x^2}{(\delta m_Q)^2} R_n] d\Phi, \label{ddz}
\end{equation}
where
\begin{eqnarray}
R_s &=& p\cdot p_i ~ p_{\ell} \cdot p_f - m_i p\cdot s ~ p_{\ell}\cdot p_f, \label{rs} 
~~~~~ \ ~~~~~ \ ~~~~~ \ 
\\
R_i^e &=& m_f(1+\rho)(p\cdot p_i ~ s\cdot q - q\cdot p_i ~ s\cdot p + m_i p \cdot q) \nonumber
\\
&+&(1-\rho)[q\cdot p_i ~ p_f\cdot p + q\cdot p_f ~ p_i\cdot p - p_i \cdot p_f ~ q \cdot p \nonumber
\\
&+& m_i (s\cdot p_f ~p\cdot q - s\cdot p ~ p_f\cdot q ~ - s\cdot q ~ p_f\cdot p)], \label{re}  ~~~~~ \ ~~~~~ \ ~~~~
\\
R_i^o &=& -[m_i(1-\rho)-m_f(1+\rho)]
\epsilon_{\alpha\beta\mu\nu} s^{\alpha}p_f^{\beta}p_i^{\mu}p^{\nu}, \label{toas} 
~~~~~ \ ~~~~~ \ ~~~~~ \ ~~~~~ \
\\
R_n &=& p\cdot p_{\ell} ~ [(2 q\cdot p_i ~ q\cdot p_f - q^2 p_f\cdot p_i)(1+\rho^2) \nonumber
\\
&+&2m_i \rho (2 q\cdot p_f ~ s\cdot q - q^2 p_f\cdot s)+ m_i m_f (1-\rho^2)q^2]. \label{rn} 
\end{eqnarray}

\section{Partial Decay Width}

In order to determine this observable, we have to integrate the expression (\ref{ddz}) over the whole phase space and to average over the spin states
of the initial baryon. To this end, we adopt a frame where the parent resonance is at rest. The calculation, whose details are given in the Appendix, yields
\begin{equation}
\Gamma = \frac{|V_{cb}|^2 G^2}{2^7\pi^3 m_i} \int_{m_f}^{E_f^m} dE_f \zeta_0^2(q^2)
\int_{E_{\ell}^-}^{E_{\ell}^+} dE_{\ell} J_0(E_f,E_{\ell}). \label{das}
\end{equation}
Here
\begin{equation} 
J_0(E_f,E_{\ell}) = 2^7 {\bar R}_s + 2^5 x \frac{m_{\ell}}{\delta m_Q} {\bar R}_i^e cos\varphi + 2^4 \frac{x^2}{(\delta m_Q)^2} {\bar R}_n,
\end{equation}
where the barred quantities are obtained from eqs. (\ref{rs}), (\ref{re}) and (\ref{rn}) by taking into account four-momentum conservation and spin average, as illustrated in the Appendix: they depend 
just on the energies $E_f$ and $E_{\ell}$, respectively, of the final hadron and of the charged lepton in the $\Lambda_b$ rest frame.
The limits of integration are deduced in the Appendix, using four-momentum conservation. The partial width (\ref{das}) assumes the form
\begin{equation} 
\Gamma(x;\rho;\varphi) = \Gamma^S 
[1+xcos\varphi(\alpha_s-\alpha_p\rho)+x^2(\beta_s+\beta_p\rho^2)], \label{gtau}
\end{equation}
where $\Gamma^S $ is the SM prediction for the partial width of the decay and the coefficients $\alpha_{s(p)}$ and $\beta_{s(p)}$ characterize the NP contribution. As regards the IW form factor, we adopt the one inferred by ref. \cite{klw}, 
{\it i. e.},
\begin{equation} 
\zeta_0 (q^2) = \tilde{\zeta}_0 [\omega(q^2)] = 
1 -1.47(\omega-1)+0.95(\omega-1)^2, \ ~~~~ ~~~~ \ \omega = \frac{m_i^2+m_f^2-q^2}{2m_im_f}.
\end{equation} 
We consider also, as an alternative, one of the hadronic currents employed by ref. \cite{swd}, which we denote as SR, because it is deduced from the sum rules. A sketch of the relative calculation is given in the Appendix. 

The numerical inputs are\cite{pdg} $G$ = $1.166379\cdot 10^{-5}$ $GeV^{-2}$, $m_i$ = $5.619$ $GeV$, $m_f$ = $2.286$ $GeV$ and $ V_{cb}$ = 0.0411. 

For $\ell$ = $\tau$, setting $m_{\ell}$ = $1.777$ $GeV$, the numerical integration yields 
\begin{equation} 
\Gamma_{\tau}^S = 5.633 ~ (4.280)\cdot 10^{-15} GeV \label{wtau}
\end{equation}
and 
\begin{equation} 
\alpha_s=0.453 ~ (0.270), ~~~~\alpha_p=0.092~ (0.082), ~~~~ 
\beta_s=0.402 ~ (0.751), ~~~~ \beta_p=0.078 ~ (0.050). 
\end{equation}The values inside the parentheses refer, here and in the following, to the predictions relative to the SR assumption for the hadronic current.

For a light lepton, we set $m_{\ell}$ $\to$ 0, so that $\alpha_s = \alpha_p$ = 0; moreover
\begin{equation} 
\Gamma_{\ell}^S = 3.300 ~ (2.067)\cdot 10^{-14} GeV, ~~~~ \beta_s=0.566 ~ (0.359), ~~~~ \beta_p=0.235 ~ (0.089) \label{wlp}. 
\end{equation}
Insertion of the mass of the muon does not substantially change the results.

The experimental value of the partial width is 
\begin{equation} 
\Gamma_{\ell}^{exp} = (2.95_{-1.14}^{+1.45})\cdot 10^{-14} 
GeV\cite{pdg}. \label{exp}
\end{equation}
The ratio $\Gamma_{\tau}^S/\Gamma_{\ell}^S$ is 0.171 (0.207), practically model independent, but significantly different from the one found by other authors\cite{swd}.

We consider the observable 
\begin{equation} 
R = \frac{\Gamma-\Gamma^S}{\Gamma^S} \label{rat1} 
\end{equation}
and exhibit in fig. 1, both for the heavy lepton and for a light lepton, the behavior of $R$ as a function of $x$ ($z$) and of $\varphi$, in the case of only (pseudo-)scalar coupling. 

\begin{figure}[h]
\centering
\includegraphics[width=1.00\textwidth] {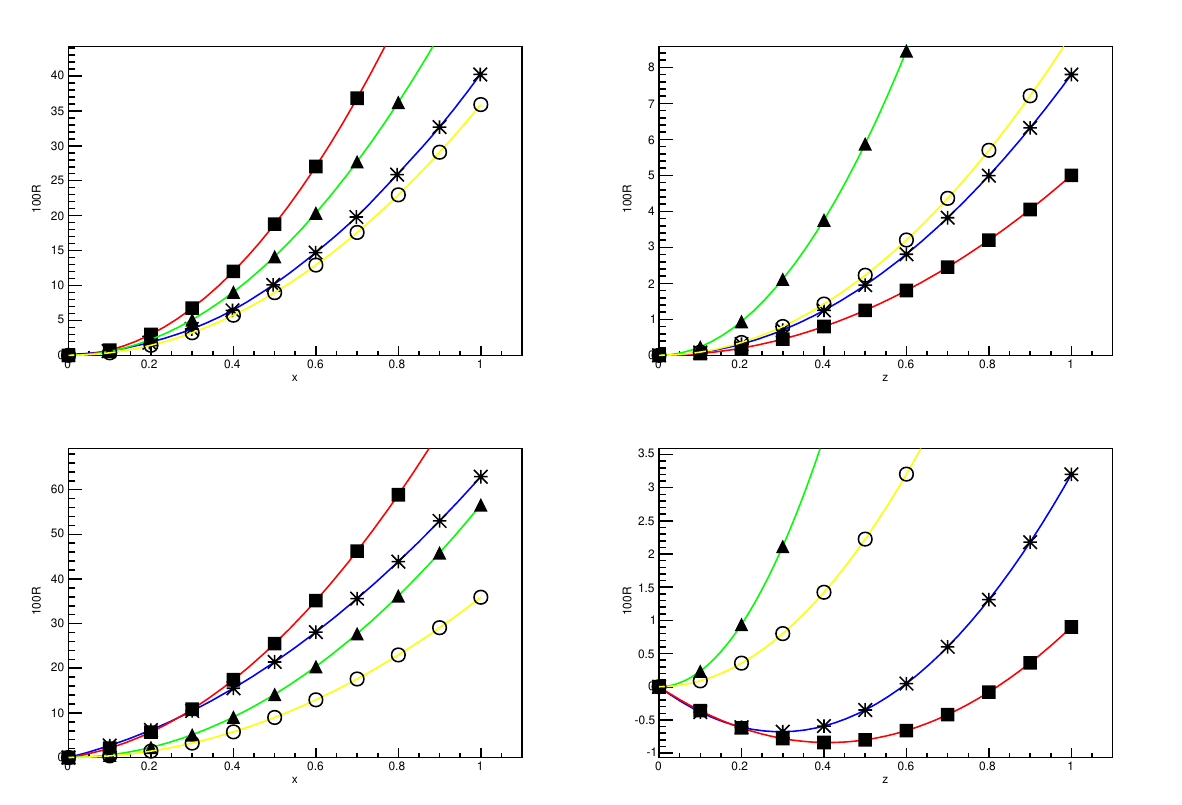}
\caption{The ratio (\ref{rat1}) as a function of $x(z)$, both for the $\tau$ (stars for IW, squares for SR) and for the light leptons (triangles for IW, circles for SR). Upper panels:  $\varphi$ = $\pi/2$; in the $1^{st}$ ($2^{nd}$) one, only the (pseudo-)scalar coupling is considered. Lower panels: the same, with $\varphi$ = $\pi/3$. \label{fig:Gamma}}  
\end{figure}
We show in Table 1 the values of $x$ ($z$) allowed according to the analysis of ref. \cite{swd} in the (pseudo)-scalar case, and the corresponding values of $R$: our choice to limit ourselves 
to $|\varphi|$ $\leq$  $\pi/2$ entails to consider, for the scalar coupling, only the boundary value of $x$ for any $\varphi$, while, in the pseudo-scalar case, an interval of $z$-values is taken into account.  

We analyze also the case of a charged higgs exchange, eq. (\ref{2hdm}). For $\ell$ = $\tau$, the ratio (\ref{rat1}) remains sizable down to $x$ $\sim$ 0.2, provided $|\varphi|$ is less than $\sim$ $\pi/3$. With a light lepton, an appreciable value of $R$ is obtained for $x$ $\geq$ $0.35 ~ (0.25)$. In this connection, we note that the IW predictions are slightly smaller than those by SR if a light lepton is detected, while the situation is inverted for the $\tau$ lepton.

\begin{table*}
\begin{center}
\caption{The ratio (\ref{rat1}) for the $\tau$ decay, in the ranges allowed by ref. \cite{swd}, $|\varphi|$ $\leq$ $\pi/2$: scalar ($S$, $x$) and pseudo-scalar ($P$, $z$) coupling.
}
\begin{tabular}{|c|c|c|c|c|}
\hline\hline
$\varphi$&$~~x~~$&$~~R~(S)~~$&$~~z~~$&$~~R~(P)~~$ \\
\hline\hline
0.0     &  0.40  & 0.25(0.23) &  0.50 \textemdash 2.50 & -0.03(-0.03) $\to$ 0.26(0.11)\\
$\pi/6$ &  0.43  & 0.24(0.24) & 0.56 \textemdash 2.63 & -0.02(-0.02) $\to$ 0.33(0.16) \\
$\pi/3$ &  0.55  & 0.25(0.30) &  0.77 \textemdash 3.03 &
+0.01(0.00) $\to$ 0.58(0.34)\\
$\pi/2$ &  0.80  & 0.26(0.50) & 1.32 \textemdash 3.71 &+0.14(0.09) $\to$ 1.07(0.69) \\
\end{tabular}
\label{tab:one}
\end{center}
\end{table*}
\vskip 0.50cm

\section{T-odd Asymmetry} 

The observable (\ref{as}) corresponds to the ratio
\begin{equation}
{\cal A} = \frac{\Delta\Gamma}{\Gamma}, \label{asy}
\end{equation} 
where $\Delta\Gamma$ = $\Gamma^+$-$\Gamma^-$ and $\Gamma^{\pm}$ is calculated by integrating the third term of eq. (\ref{ddz}) over the half-space  where the correlation (\ref{crr}) is
positive (negative). We show in the Appendix that  
\begin{equation}
\Delta\Gamma = -\frac{xG^2}{2^3\pi^2}\frac{m_{\ell}H(\rho)}{\delta m_Q}|V_{cb}|^2 |{\bf P}| sin\varphi \int_{m_f}^{E_f^m} dE_f \zeta_0^2(q^2) \frac{1}{\sqrt{-a}}(c-\frac{b^2}{4a}). \label{dlgm}
\end{equation}
Here
\begin{eqnarray} 
H(\rho) &=& m_f-m_i+(m_f+m_i)\rho, \ ~~~~~ \ \ ~~~~~ \
\\
a &=& -q^2 = 2m_iE_f-m_i^2-m_f^2, \ ~~~~~ \ \ ~~~~~ \ 
\\
b &=& 2m_iE_f^2 (2m_i^2+M^2)E_f+M^2m_i, \ ~~~~~ \ \ ~~~~~ \ 
\\
c &=& -(m_i^2+m^2_{\ell})E_f^2+m_iM^2E_f+m_f^2 m^2_{\ell}-\frac{1}{4}M^4 \ ~~~~~ \   
\end{eqnarray}
and $M^2$ = $m_i^2+m_f^2+m_{\ell}^2$.

Performing the numerical integration in eq. (\ref{dlgm}), and then inserting this and eq. (\ref{gtau}) into eq. (\ref{asy}),
yields, for $\ell$ = $\tau$, 
\begin{equation}
{\cal A}_{\tau} = \frac{(\gamma_s-\gamma_p\rho) x sin\varphi|{\bf P}|}{1+R}, \label{asmm}
\end{equation}
with $\gamma_s$ = 0.0292 (0.0275) and $\gamma_p$ = 0.0693 (0.0328). 
The behavior of ${\cal A}_{\tau}$ is represented in Fig. 2, as well as in Table 2, assuming a polarization $|{\bf P}|$ = 1. ${\cal A}$ vanishes when a light lepton is detected, due to the factor $m_{\ell}$ in eq. (\ref{dlgm}).  
  
The asymmetry (\ref{asmm}) vanishes for $\rho$ = 0.42 (0.52). Since these values are close to the value (\ref{2hdm}), corresponding to 2HDM, in this case the asymmetry is negligibly small.

Lastly, it is straightforward to show that, in the case of NP vector or axial coupling\cite{swd,dut}, the T-odd asymmetry would vanish. In this connection, we observe that, in the helicity formalism\cite{swd}, the interference term between the SM amplitude and the NP amplitude appears explicitly only as regards the scalar and the pseudo-scalar coupling.

\begin{figure}[h]
\centering
\includegraphics[width=1.00\textwidth] {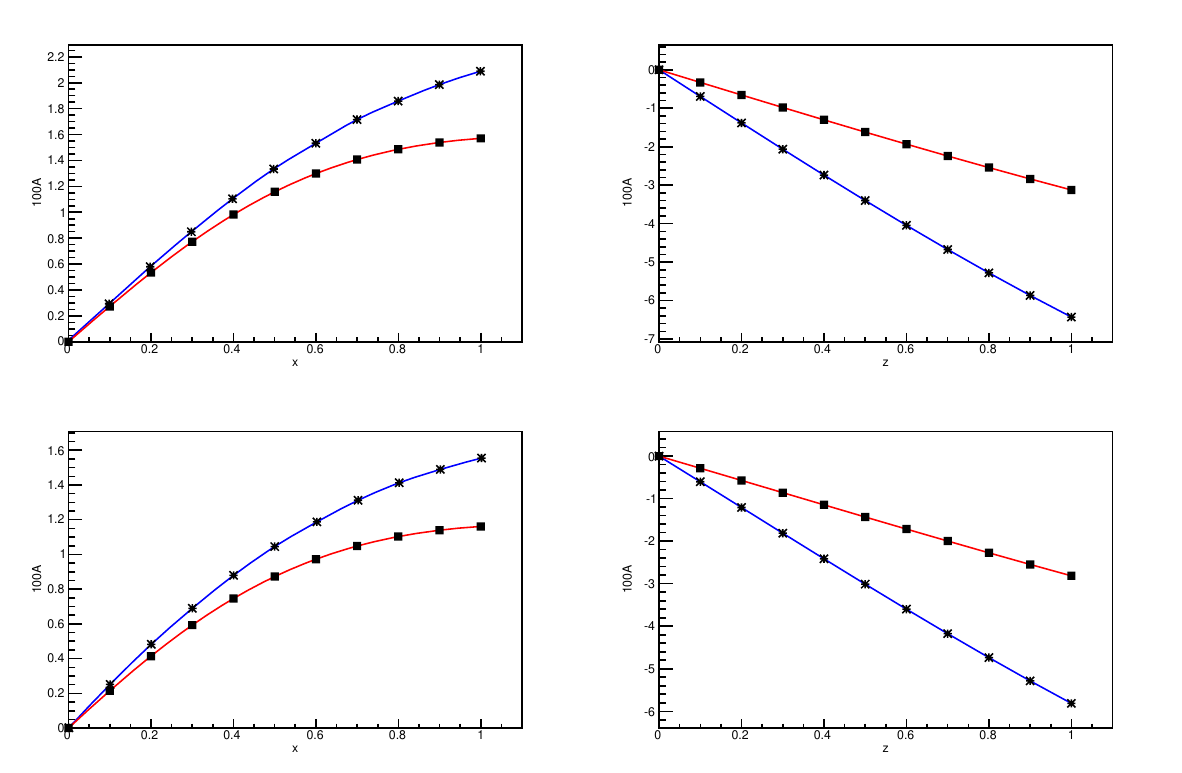}
\caption{ The T-odd asymmetry (\ref{asmm}) for the $\tau$ lepton:
stars for IW, squares. A polarization $|{\bf P}|$ = 1 
assumed. Upper and lower panels: same conditions and notations as for
fig. 1.
\label{fig:Asymm}}  
\end{figure}

\section{Conclusions}
We have studied the sensitivities to NP - precisely, to the scalar and/or pseudo-scalar coupling - of the partial widths and of the T-odd asymmetries (\ref{as}) of the decays (\ref{slh}). The sensitive quantities, denoted, respectively, as $R$, eq. (\ref{rat1}), and as ${\cal A}$, eq. (\ref{asmm}), have been expressed as functions of the real parameters $x$, $\varphi$ and $\rho$, examining three particular cases.
 
The asymmetry can be realistically detected only for
the $\tau$ decay, provided the $\Lambda_b$ is sizably polarized and the NP is described by the pseudo-scalar coupling, with a sufficiently large strength; in this connection, it is worth noting that such a condition is satisfied by the limits on NP inferred from the experiments\cite{swd}. 

On the contrary, $R$ reaches considerable values for the scalar coupling
and it is also sensitive to a charged higgs exchange, even at small $x$ $\sim$ 0.2-0.4, both for the light leptons and for the heavy one. As regards the pseudo-scalar coupling, the ratio becomes appreciable only for $z$ $\geq$ 0.5-0.7, but it is considerably larger in the case of light leptons.

Our predictions do not depend so dramatically on the model adopted;
in fact, they are sometimes hardly model dependent, as, for example, 
in the case of the ratio $R$ for the pseudo-scalar coupling with light leptons.

Our method can be extended to other semi-leptonic decays of beauty baryons.
\vskip 0.50cm
\centerline{\bf Acknowledgments}
\vskip 0.30cm
One of the authors (EDS) is grateful to his friend F. Fontanelli for useful suggestions and discussions.
\newpage

\begin{table*}
\begin{center}
\caption{The T-odd asymmetry (\ref{asmm}): same conditions and notations as for Table 1. }
\begin{tabular}{|c|c|c|c|c|}
\hline\hline
$\varphi$&$~~x~~ $&$~~{\cal A}~(S)~~$&$~~z~~ $&$~~{\cal A}~(P)~~$ \\
\hline\hline
0.0     &  0.40  & 0.00 & 0.50 \textemdash 2.50 & 0.00 \\
$\pi/6$ &  0.43  & 0.01(0.01) & 0.56 \textemdash 2.63 & -0.02(-0.01) $\to$ -0.07(-0.04)\\
$\pi/3$ &  0.55  & 0.01(0.01) & 0.77 \textemdash 3.03 & -0.04(-0.02) $\to$ -0.12(-0.06) \\
$\pi/2$ &  0.80  & 0.02(0.01) & 1.32 \textemdash 3.71 & -0.08(-0.04) $\to$ -0.12(-0.07) \\
\end{tabular}
\label{tab:two}
\end{center}
\end{table*}  

\setcounter{equation}{0}
 \renewcommand\theequation{A. \arabic{equation}}

 \appendix{\large \bf Appendix}
\vskip 0.30cm
Here we calculate the partial width and the T-odd asymmetry of the decay $\Lambda_b \to \Lambda_c \ell \bar{\nu}_{\ell}$ under the IW assumption. Moreover, we sketch the calculation of these observables with the most general semi-leptonic form factor.
\vskip 1cm
\centerline{\bf A) Partial Decay Width}
\vskip 1cm
As regards the partial width, we have to integrate the differential decay width (\ref{ddz}) over the whole phase space,
averaging over the spin states of $\Lambda_b$. We get
\begin{equation}
\Gamma = \frac{1}{2m_i} |V_{cb}|^2 \frac{G^2}{2}\int d\Phi \zeta_0^2(q^2) J({\bf p}_f,{\bf p}_{\ell}), \label{ddc}
\end{equation}
where
\begin{equation}
J({\bf p}_f,{\bf p}_{\ell}) =  2^7 R_s + 2^5 x \frac{m_{\ell}}{\delta m_Q} R_i^e cos\varphi + 2^4 \frac{x^2}{(\delta m_Q)^2} R_n
\end{equation}
and $R_s$, $R_i^e$ and $R_n$ correspond to eqs. (\ref{rs}), (\ref{re})  and (\ref{rn}), after taking spin average.
In  a reference frame at rest with respect to $\Lambda_b$, we have 
\begin{equation}
d\Phi = \frac{d^3p}{2(2\pi)^3|{\bf p}|} \frac{d^3p_f}{2(2\pi)^3E_f} \frac{d^3p_{\ell}}{2(2\pi)^3E_{\ell}}
(2\pi)^4\delta^3({\bf p}+{\bf p}_f+{\bf p}_{\ell})\delta(m_i-E_f-E_{\ell}-|{\bf p}|).
\end{equation}
Moreover 
\begin{eqnarray}
R_s &=& m_i(E_f E_l-{\bf p}_f\cdot{\bf p}_{\ell})(m_i-E_f-E_{\ell}), 
~~~~~ \ ~~~~~ \ ~~~~~ \ 
\\
R_i^e &=& m_i[(1+\rho){\cal S}_1 +(1-\rho){\cal S}_2], ~~~~~ \ ~~~~~ \ ~~~~
\\
{\cal S}_1 &=& m_f [m_i(m_i-2E_f-E_{\ell}) + m_f^2 + E_f E_{\ell} -{\bf p}_f\cdot{\bf p}_{\ell}],
\\
{\cal S}_2 &=& (m_i-E_f)(m_i E_f - E_f E_{\ell} - m^2_f + {\bf p}_f\cdot{\bf p}_{\ell}) \nonumber
\\ 
&+& (m_i-E_f-E_{\ell})(m_iE_f-m_f^2) \nonumber ~~~~~ \ ~~~~~ \ ~~~~~ \ ~~~~~
\\
&-& E_f[m_i(m_i-2E_f-E_{\ell})+ m_f^2+ E_f E_{\ell} -{\bf p}_f\cdot{\bf p}_{\ell}],
\\
R_n &=& m_i[E_{\ell} (m_i-E_f)-m_{\ell}^2+{\bf p}_f\cdot{\bf p}_{\ell}]\{(1+\rho^2)
\nonumber 
\\
&\times& [2(m_i-E_f)(m_iE_f-m_f^2)-E_f(m_i^2+m_f^2-2m_iE_f)] \nonumber 
\\
&+&(1-\rho^2)m_f(m_i^2+m_f^2-2m_iE_f)\}.
\end{eqnarray}
Lastly, four-momentum conservation entails
\begin{eqnarray}
2{\bf p}_f\cdot{\bf p}_{\ell} &=& M^2-2m_i(E_f+E_{\ell})+2E_fE_{\ell}, \ ~~~~~ \  \label{cns1}
\\
{\bf p}^2 &=& {\bf p}_f^2 + {\bf p}_{\ell}^2 +2{\bf p}_f \cdot {\bf p}_{\ell}, \ ~~~~~ \ \ ~~~~~ \ 
\ ~~~~~ \ 
\end{eqnarray}
with
\begin{equation}
M^2 = m_i^2+m_f^2+m_{\ell}^2,
\end{equation}
having used the same notations as in the text for the various particles involved. Note that the scalar product ${\bf p}_f\cdot{\bf p}_{\ell}$ is just a function of $E_f$ and $E_{\ell}$; therefore, taking account 
of the expressions $R_s$, $R_i^e$ and $R_n$, we shall substitute, from now on,
\begin{equation}
J({\bf p}_f,{\bf p}_{\ell}) \to J_0(E_f,E_{\ell}). \label{rr0}
\end{equation}
Setting $\varepsilon$ = $E_f+E_{\ell}+|{\bf p}|-m_i$, and defining $\theta$ as the angle between ${\bf p}_f$ and ${\bf p}_{\ell}$ and $\phi$ as the azimuthal angle of ${\bf p}_{\ell}$, we get 
\begin{equation}
d^3p_{\ell} ~ \delta({\varepsilon}) = {\bf p}_{\ell}^2 dp_{\ell} ~ d\phi ~ d cos\theta ~ \delta({\varepsilon}) =
\frac{|{\bf p}_{\ell}||{\bf p}|}{|{\bf p}_f|} ~ dp_{\ell} ~ 
d\phi ~ d\varepsilon ~ \delta({\varepsilon}); \label{rel1} 
\end{equation}
here we have exploited the fact that $d\varepsilon/d cos\theta$ = $d|{\bf p}|/d cos\theta$ = 
$|{\bf p }_f||{\bf p}_{\ell}|/|{\bf p}|$.  

Moreover, the relation $E_f$ = $\sqrt{m_f^2+{\bf p}_f^2}$ implies 
\begin{equation}
dE_f = \frac{|{\bf p}_f|dp_f}{E_f}, \label{rel2} 
\end{equation}
an analogous equation holding for $d E_{\ell}$. Inserting eqs. (\ref{rr0}) to (\ref{rel2}) into the integral (\ref{ddc}), and taking account that $q^2$ = $m_i^2+m_f^2-2m_iE_f$, yields 
\begin{equation}
\Gamma = \frac{|V_{cb}|^2 G^2}{2^7\pi^3 m_i} \int_{m_f}^{E_f^m} dE_f \zeta_0^2(q^2)
\int_{E_{\ell}^-}^{E_{\ell}^+} dE_{\ell} J_0(E_f,E_{\ell}). \label{dass}
\end{equation}

The limits of integration for $E_{\ell}$ are deduced from eq. (\ref{cns1}), setting 
$cos \theta$ $=$ $\pm 1$:
\begin{eqnarray}
E_{\ell}^{\pm}&=& \frac{-b\pm\sqrt{\Delta}}{2a},  \ ~~~~~ \ \ ~~~~~ \ 
\Delta = b^2-4ac, \label{zrs}
\\
a &=& -q^2 = 2m_iE_f-m_i^2-m_f^2, \ ~~~~~ \ \ ~~~~~ \ \label{coef1}
\\
b &=& 2m_iE_f^2 - (2m_i^2+M^2)E_f+M^2m_i, \ ~~~~~ \ \ ~~~~~ \ 
\\
c &=& -(m_i^2+m^2_{\ell})E_f^2+m_iM^2E_f+m_f^2 m^2_{\ell}-\frac{1}{4}M^4. \label{coef2} \ ~~~~~ \   
\end{eqnarray}
As regards the upper limit $E_f^m$, imposing ${\bf p}_{\ell}$ = 0 yields
\begin{equation}
E_f^m = \sqrt{m_f^2+ p_m^2}, \ ~~~~~ \ \ ~~~~~ \  p_m = 
\frac{1}{2}(m_i-m_{\ell}-\frac{m_f^2}{m_i-m_{\ell}}).
\end{equation}

\centerline{\bf B) T-odd Asymmetry}
\vskip 0.50cm
This asymmetry is defined by the fraction (\ref{asy}) in the text. The denominator has been calculated in the previous section of this Appendix. In order to calculate the numerator, we have to fix the azimuthal plane for ${\bf p}_{\ell}$ in the rest frame of $\Lambda_b$. To this end, we choose the $z$-axis along 
${\bf p}_f$ and define the azimuthal plane as the one singled out by the vectors ${\bf p}_f$ and ${\bf p}_f\times{\bf P}$, 
${\bf P}$ being the polarization vector of the parent resonance. Recalling eq. (\ref{toas}), we have, according to the convention (\ref{crr}),
\begin{eqnarray}
\Delta\Gamma &=& \frac{1}{2m_i} |V_{cb}|^2 \frac{G^2}{2} 2^5 x \frac{m_{\ell}}{\delta m_Q}
sin\varphi\int d\Phi' [\int_{-\pi/2}^{\pi/2}-\int_{\pi/2}^{3\pi/2}] d\phi ~ \zeta_0^2(q^2) R_i^o, \ ~~~~~ \ 
\\
R_i^o &=& H(\rho)\epsilon_{\alpha\beta\mu\nu}s^{\alpha}p_f^{\beta}p_i^{\mu}p^{\nu} =
H(\rho) m_i ~ {\bf p}_{\ell}\times {\bf p}_f\cdot{\bf P}, \ ~~~~~ \ \ ~~~~~ \ 
\ ~~~~~ \ 
\\
H(\rho) &=& m_f(1+\rho)-m_i(1-\rho). \ ~~~~~ \ \ ~~~~~ \ \ ~~~~~ \ \ ~~~~~ \
\ ~~~~~ \   
\end{eqnarray}
$d\Phi'$ is defined in such a way that $d\Phi' d\phi$ = $d\Phi$. In the frame defined above, it results
\begin{equation}
R_i^o = -m_i H(\rho)|{\bf P}||{\bf p}_{\ell}||{\bf p}_f| sin\theta_f sin\theta cos\phi, 
\end{equation}
$\theta_f$ being the angle between ${\bf P}$ and ${\bf p}_f$. Moreover, eq. (\ref{cns1}) implies
\begin{equation}
cos\theta = \frac{N(E_f, E_{\ell})}{2|{\bf p}_f||{\bf p}_{\ell}|},  \ ~~~~~ \ \ ~~~~~ \
N(E_f, E_{\ell}) = M^2 -2m_i(E_f+E_{\ell})+2E_f E_{\ell}.  
\end{equation}
Proceeding analogously to the previous integration, we get
\begin{equation}
\Delta\Gamma = -\frac{xG^2}{2^4\pi^3}\frac{m_{\ell}H(\rho)}{\delta m_Q}|V_{cb}|^2 |{\bf P}| sin\varphi \int_{m_f}^{E_f^m} dE_f \zeta_0^2(q^2) \int_{E_{\ell}^-}^{E_{\ell}^+} dE_{\ell} 
(a E_{\ell}^2+b E_{\ell}+c)^{1/2},
\end{equation}
$a$, $b$ and $c$ being given by eqs. (\ref{coef1}) to (\ref{coef2}). The integration over 
$E_{\ell}$ - performed between the two zeroes of the integrand, according to eqs. (\ref{zrs}) - yields
\begin{equation}
\Delta\Gamma = -\frac{xG^2}{2^4\pi^2}\frac{m_{\ell}H(\rho)}{\delta m_Q}|V_{cb}|^2 |{\bf P}| sin\varphi \int_{m_f}^{E_f^m} dE_f \zeta_0^2(q^2) \frac{1}{\sqrt{-a}}(c-\frac{b^2}{4a}). 
\end{equation}

\centerline{\bf C) Use of General Form Factors}
\vskip 0.50cm
The most general matrix element for the decay considered reads as
\begin{equation}
{\cal M} = V_{cb} (h_s {\tilde J}_{\mu} j^{\mu}+ 
e^{i\varphi} h_n {\tilde J}j).
\end{equation}
Here $h_s$, $h_n$, $j^{\mu}$ and $j$ are given by eqs. (\ref{cnst}) and (\ref{lcur}), while
\begin{equation}
{\tilde J}_{\mu} = V_{\mu}-A_{\mu}, \ ~~~~~ \ 
{\tilde J}=\frac{q^{\alpha}}{\delta m_Q} (V_{\alpha}+\rho A_{\alpha})
\end{equation}
and 
\begin{equation}
V_{\mu} = {\bar u}_f (X_0\gamma_{\mu}+A P_{+\mu}+A'q_{\mu}) u_i, 
\ ~~~~~ \  
A_{\mu} = {\bar u}_f (Y_0\gamma_{\mu}+B P_{+\mu}+B'q_{\mu})\gamma_5 u_i.
\end{equation}
Here $X_0$, $Y_0$, $A$, $B$, $A'$ and $B'$ are functions of $q^2$ and $P_+$ = $p_i+p_f$. Use has been made of the equations of motion. For the form factors adopted by ref. \cite{swd}, we have $A'$ = $B'$ = 0 and
\begin{equation}
X_0=f_1-(m_i+m_f)f_2, \ ~~~~~ \ Y_0= f_1+(m_i-m_f)f_2, ~~~~~ A=B= f_2,
\end{equation}
where $f_1$ and $f_2$ are the form factors listed in Table 1 of that paper. In particular, in the present letter, we adopt the form factors $f_1$ = $6.66/(20.27-q^2)$ and $f_2$ = $-0.21/(15.15-q^2)$\cite{swd}. 
The calculations of the observables proceed along the line traced in our text.

\end{document}